\documentclass[proof]{pasj00}
\draft

\begin{document}
\SetRunningHead{Author(s) in page-head}{Non-LTE Decretion Disk}
\Received{2007/11/06}
\Accepted{2008/03/25}

\title{Non-LTE Calculation for the Be Star Decretion Disk}

\author{Hidetoshi \textsc{Iwamatsu} and Ryuko \textsc{Hirata}}
\affil{Department of Astronomy, Faculty of Science, Kyoto University, Sakyo-ku, Kyoto, 606-8502}
\email{iwamatsu@kusastro.kyoto-u.ac.jp, hirata@kusastro.kyoto-u.ac.jp}


\maketitle

\begin{abstract}
The non-LTE state of the hydrogen gas in isothermal transonic decretion disks
around a B1V star has been calculated by an iterative method in order to explore
the basic physical process in the disk. This dynamical model is characterized by
a density law in the equatorial plane of $\rho(R) \propto R^{-3.5}$. The continuous
radiation is calculated with the $\Lambda$ iteration in the integral form, while
we adopt a single-flight escape probability for lines. We describe the non-LTE
state, the radiation flow and conversion in the disk. We conclude that the stellar
Balmer continuum plays a key role in the non-LTE state of the disk. The examination
of the local energy gain and loss suggests that the disk temperature has double
minima along the equatorial plane in the optically thick case: the intermediate
region caused by deficient ultraviolet radiation and the outer Lyman\ $\alpha$
cooling region. We have also calculated some observable quantities, such as the
spectral energy distribution, the $UBV$ colors, the infrared excess and the Balmer
line profiles. Our calculations with the mass loss rate less than $10^{-10}
M_{\odot}\ {\rm yr}^{-1}$ reproduce the observed continuum quantities. However,
we could not get large H$\alpha$ emission strength observed in Be stars. We
suggest that the density gradient of the Be star disk is slower than that of the
isothermal decretion disk.
\end{abstract}

\section{Introduction}\label{Introduction}
Be stars are classified as main sequence or giant B stars which have ever shown
the emission in Balmer lines. \citet{Struve1931}, in his pioneering work on Be
stars, proposed flattened equatorial disks around Be stars from the correlation
between the emission line widths and the projected rotational velocities of 
underlying stars. \citet{Coyne1969} supported this view from the polarimetric
observations. While, the rotation law in the disk had been unsettled for a long
time. It seems that it has been recently established that the disk is rotating
with a Keplerian law, because the one-armed oscillation, which is excited only
in the Keplerian or the near Keplerian disk, can explain the observational facts
such as the V/R variation \citep{Okazaki1997} and the Balmer progression
\citep{Hirata2000}. The most confident evidence for the Keplerian disk is the
interferometric detection of a size difference of the H$\alpha$ emission region
in the anti-direction measured from the central star, which correlates with the
V/R variation (\authorcite{Vakili1998} \yearcite{Vakili1998}; \authorcite{Berio1999}
\yearcite{Berio1999}). \citet{Hummel2000} obtained the same conclusion from the
comparison between the observed and calculated emission profiles. The decretion
disk, which was introduced first by \citet{Lee1991} and later developed by
\citet{Porter1999} and \citet{Okazaki2001}, gives the physical basis of the
Keplerian disk, though the mechanism which supplies the mass and angular momentum
is not yet known. The Be star disk model with magnetic field has been devised by
\citet{Cassinelli2002}, \citet{Maheswaran2003} and \citet{udDoula2002}. However,
\citet{Owocki2003} concluded that the simulation with the stellar rotation does
not form a magnetically torqued disk with enough density. 

It is well known that the Be star disk is highly in non-LTE state, because the
density is not so high that the collisional process is effective. It is necessary
to solve non-LTE problem to derive physical quantities from observational materials.
Then, it is highly desirable to calculate non-LTE state for the decretion disk
model. Several non-LTE calculations have been published for the flattened disks
with different velocity structures, namely, different density structures.
\citet{Marl1969} and \citet{Marl1978} adopted the weighted mean of some extreme
cases as the non-LTE solution. \citet{Millar1998} and \citet{Millar1999} expanded
the model to the radiative equilibrium model. The latter took into account the
diffuse radiation in an approximate way. The model has been developed further by
\citet{Jones2004} and \citet{Sigut2007}, in which they examined the metal effect
to the thermal structure of the disk. \citet{Sigut2007} adopted a single-flight
escape probability in the static plane-parallel geometry for lines and on-the-spot
approximation for continuous radiation. They also took into account the diffuse
continuous radiation in an approximate way. Non-LTE calculations were also performed
by \citet{Kriz1979}, \citet{Stee1995}, and \citet{Stee2001}, based on Sobolev
approximation. This approximation is an adequate one when the velocity gradient is
large (\cite{Hamann1981}; \cite{deKoter1993}). However, the expansion velocity in
the decretion disk is very small \citep{Okazaki1997}. \citet{Hummel1994} and
\citet{Hummel1997} adopted the equivalent two level approximation and calculated
H$\alpha$ line profiles, using $\Lambda$ iteration. \citet{Carciofi2006} performed
the non-LTE calculation for the decretion disk under the radiative equilibrium by
using Monte Carlo method. \citet{Carciofi2007} improved their model to take into
account the vertical temperature structure. They showed that there is a temperature
minimum in the several stellar radii on the equatorial plane.

In this paper, we present our results based on the full treatment of the diffuse
continuous radiation, as an extension of \citet{Iwamatsu2007}. In section
\ref{Model}, we describe the central star and the disk structure for which we
calculate the non-LTE state. In section \ref{Calculation}, we describe our method
of the non-LTE calculation. Section \ref{NonLTEstate} gives the non-LTE state in
our model, the local energy balance analysis, and also the total energy gain and
loss in the whole disk. In section \ref{ContProperty}, we describe the continuum
properties such as the spectral energy distributions, the $UBV$ colors, and the
infrared excess.  Section \ref{LineProperty} treats the line properties such as
the Balmer line profiles, the equivalent widths, and the Balmer decrement. The
summary and the concluding remarks are given in section \ref{Conclusion}.

\bigskip

\section{Central Star and Decretion Disk}\label{Model}
The spectral type of the central star is assumed to be B1V and its basic parameters
are chosen as the effective temperature $T_{\rm eff}=24000$\,K, the radius
$R_*=6.9\,\RO$, and the mass $M_*=11.0\,\MO$, referred to $\pi$ Aqr
\citep{Bjorkman2000}. The stellar flux is taken from ATLAS9 \citep{Kurucz1993}
with the other parameters: the gravitational acceleration at the stellar surface,
$\log g=4.0$, the solar abundance, and the turbulent velocity of 2~km\,s$^{-1}$. 
For easier integration over the frequency, we adopt the stellar flux without line
blanketting. Although the effective temperature for the stellar flux without 
line blanketting is actually 26400\,K, we expect that the influence of line
blanketting is small, because the flux difference near each shortward continuum 
limit is small. We assume the central star is spherical without limb darkening. 

As a dynamical model, we adopt an isothermal transonic decretion disk in steady
state with a viscosity parameter $\alpha=0.1$ \citep{Okazaki2001}. This
one-dimensional model is characterized by almost Keplerian rotation and small
expansion velocity. We assume that the disk is axisymmetric about the rotational
axis and symmetric about the equatorial plane, and is in hydrostatic equilibrium
in the vertical direction. It is convenient to adopt a cylindrical coordinate,
whose origin and the polar axis are the center of the star and the rotation axis
of the star, respectively. Let us denote the distance from the rotation axis by
$\varpi$ and the vertical distance from the equatorial plane by $z$ in units of
the stellar radius. We assume that the expansion and rotational velocities at any
point are the same as those at $\varpi$ in the equatorial plane. The number density
$N(\varpi,z)$ is related to the number density along the equatorial plane
$N(\varpi,0)$ \citep{Marl1969} as
\begin{eqnarray}\label{hydrostatic}
N(\varpi,z)=N(\varpi,0)\exp\left\{-\frac{1}{Q}
\left(\frac{1}{\varpi}-\frac{1}{\sqrt{\varpi^2+z^2}}\right)\right\}\,,
\end{eqnarray}
with 
\begin{equation}
 Q=\frac{kT_{\rm disk}R_*}{Gm_{\rm H}\mu M_*}\,,
\end{equation}
where $\mu$ is the mean molecular weight, which is fixed to 0.5 in our calculations.
$M_*$ is the stellar mass and $T_{\rm disk}$ is the temperature of the isothermal
disk. The other notations have usual meanings. The number density $N(\varpi,0)$ is
obtained from the column density of \citet{Okazaki2001}'s model under the assumption
that the disk is geometrically thin. This assumption is given by 
\begin{equation}\label{approximation}
 -\frac{1}{Q}\bigg(\frac{1}{\varpi}-\frac{1}{\sqrt{\varpi^2+z^2}}\bigg)\simeq
 -\bigg(\frac{z}{\sqrt{2Q\varpi^3}}\bigg)^2.
\end{equation}
This dynamical model is characterized by $N(\varpi,0) \propto \varpi^{-3.5}$. The
number density $N(\varpi,z)$ is derived from equation (\ref{hydrostatic}). Table
\ref{OkazakiList} shows the mass loss rate, $\dot{M}$, the total mass of the disk
$M_{\rm disk}$, the assumed disk temperature $T_{\rm disk}$, and the density
$\rho(\varpi,0)$ at $\varpi=1$ and 1000 in our nine models
\footnote{In conversion from the column density to the number density in
\citet{Iwamatsu2007}, there was an error that the column density is divided by
$\mu=0.5$. Therefore, the correct density is a half of the tabulated one, and all
of the mass loss rate appeared in \citet{Iwamatsu2007} should be corrected by
multiplying two.}. The number density $N(1,0)$ is given by 
$N(1,0)=\rho(1,0)/m_{\rm H}=5.9754\times 10^{23}\rho(1,0)$.

\begin{table*}
 \begin{center}
 \caption{Physical quantities of our dynamical model sequence. We show the mass loss
          rate $\dot{M}$, the total mass of the disk $M_{\rm disk}$, the disk
          temperature $T_{\rm disk}$, and the density $\rho(\varpi,0)$, at $\varpi=1$
          and 1000. $a(b)$ means $a \times 10^{b}$.}
 \label{OkazakiList}
  \begin{tabular}{p{4em}ccccc}
   \hline
   \multicolumn{1}{c}{model}  & $\dot{M}$ & $M_{\rm disk}$ & $T_{\rm disk}$
 & \multicolumn{2}{c}{$\rho(\varpi,0)$ [${\rm g/cm}^{-3}$]} \\ \cline{5-6}
   \multicolumn{1}{c}{ }  & [$\MO\,{\rm yr}^{-1}$] & \,[$\MO$] & \,[K] &
               $\varpi=1$ & $\varpi=10^3$ \\
   \hline
    1 \dotfill & $1.0(-12)$ & $1.02(-10)$ & 1.6(4) & $1.75(-12)$ & $8.60(-24)$\\
    2 \dotfill & $2.0(-12)$ & $2.04(-10)$ & 1.6(4) & $3.50(-12)$ & $1.72(-23)$\\
    3 \dotfill & $5.0(-12)$ & $5.11(-10)$ & 1.6(4) & $8.74(-12)$ & $4.30(-23)$\\
    4 \dotfill & $1.0(-11)$ & $1.02(-09)$ & 1.6(4) & $1.75(-11)$ & $8.60(-23)$\\
    5 \dotfill & $2.0(-11)$ & $2.04(-09)$ & 1.6(4) & $3.50(-11)$ & $1.72(-22)$\\
    6 \dotfill & $5.0(-11)$ & $5.11(-09)$ & 1.6(4) & $8.74(-11)$ & $4.30(-22)$\\
    7 \dotfill & $1.0(-10)$ & $1.02(-08)$ & 1.6(4) & $1.75(-10)$ & $8.60(-22)$\\
    8 \dotfill & $2.0(-10)$ & $2.04(-08)$ & 1.6(4) & $3.50(-10)$ & $1.72(-21)$\\
    9 \dotfill & $1.0(-10)$ & $1.68(-08)$ & 1.2(4) & $3.15(-10)$ & $1.35(-21)$\\
   \hline
  \end{tabular}
 \end{center}
\end{table*}

The disk temperature in models 1$-$8 is fixed to two thirds of the effective 
temperature of the central star \citep{Hummel2000}, i.e., 16000\,K, while different
mass loss rates are assigned from $1.0\times10^{-12}$ to
$2.0\times10^{-10} \MO\ {\rm yr}^{-1}$ for models 1$-$8 in this order. Model 9 has
the same mass loss rate as model 7, but with the disk temperature of 12000\,K. This
case is introduced for the examination of the energy balance. For models 1$-$8, we
obtain the rotational velocities of $v_\phi(\varpi=1)=590~{\rm km~s}^{-1}$ and
$v_\phi(\varpi=1000)=5.35~{\rm km~s}^{-1}$ and the expansion velocities of
$v_\varpi(\varpi=1)=0.00472~{\rm km~s}^{-1}$ and
$v_\varpi(\varpi=1000)=30.6~{\rm km~s}^{-1}$. For model 9, we obtain
$v_\phi(\varpi=1)=591~{\rm km~s}^{-1}$, $v_\phi(\varpi=1000)=6.30~{\rm km~s}^{-1}$, 
$v_\varpi(\varpi=1)=0.00303~{\rm km~s}^{-1}$, and
$v_\varpi(\varpi=1000)=22.4~{\rm km~s}^{-1}$. Hence, the rotational velocities at
$\varpi=1000$ are about one third of the Keplerian velocity, while the Mach numbers
at $\varpi=1000$ are about two. In figure \ref{Isodensity}, we show the iso-contour
of the number density on the meridional plane for our model 7, together with the
disk boundary which will be described in section \ref{grids}.

\begin{figure}
 \caption{Iso-contour of the number density on the meridional plane in model 7. 
          The outer boundary is also shown by a solid curve.}
 \label{Isodensity}
\end{figure}

\bigskip

\section{Non-LTE Calculation}\label{Calculation}
\subsection{Statistical Equilibrium Equation}\label{IonExEq}
We assume that the disk consists of only hydrogen atoms. When we designate the
uppermost bound level of hydrogen atom by $n_0$, the charge conservation is
expressed as
\begin{equation}
 \label{NeDerive}
 N_e=N_+=N(\varpi,z)-\sum_{n=1}^{n_0}N_n\ ,
\end{equation}
where $N_e, N_+$, and $N_n$ are the electron number density, the proton number
density, and the population of the $n$-th level at the point $(\varpi,z)$,
respectively. The $2s$ and $2p$ sublevels are treated separately as in \citet{Marl1969}. 
We assume complete redistribution for the line radiation. Following the notation
of \citet{Kriz1974}, the statistical equilibrium equations are expressed as
\begin{eqnarray}\label{ioniexcieqO}
 & N_e\sum_{n'\neq n}N_{n'}C_{n'n}-N_nN_e\sum_{n'\neq n}C_{nn'}
  +N_e^2N_{+}C_{cn}-N_eN_nC_{nc} \nonumber & \\
 & +\sum_{n'=n+1}^{n_0}(N_{n'}A_{n'n}+N_{n'}B_{n'n}J_{nn'}-N_nB_{nn'}J_{nn'}) & \nonumber \\
 & -\sum_{n'=1}^{n-1}(N_{n}A_{nn'}+N_{n}B_{nn'}J_{n'n}-N_{n'}B_{n'n}J_{n'n})
   +N_eN_{+}\alpha_{n} & \nonumber \\
 & -4\pi\int_{\nu_n}^{\infty} a_n(\nu)\frac{J_{\nu c}}{h\nu}
   \Big\{1-\frac{1}{b_n}\exp\Big(-\frac{h\nu}{kT}\Big)\Big\} d\nu=0
   \qquad (n=1,2,\cdots,n_0)\ ,&
\end{eqnarray}
where $J_{\nu c}$, $J_{nn'}$ are the mean intensity of the continuum and line
radiation, respectively. $A_{n'n}$, $B_{nn'}$, and $B_{n'n}\ (n'>n)$ are Einstein
coefficients for the spontaneous emission, absorption, and stimulated emission,
respectively. $C_{nn'}$ is the collision coefficient, and $a_n(\nu)$, $\alpha_n$
are the ionization and recombination coefficients, respectively. Here, we neglect
two-photon emission between 1s and 2s levels. Introducing a departure factor
$b_n=N_n/N_n^{*}$, where $N_n^{*}$ is the LTE value, equations (\ref{ioniexcieqO})
are written as
\begin{eqnarray}\label{ioniexcieq}
& \sum_{n'=1}^{n-1}b_{n'}g_{n'}e^{x_{n'}}(N_eC_{n'n})
      -b_n\Big[\sum_{n'=1}^{n-1}g_{n'}e^{x_{n'}}(N_eC_{n'n}) \nonumber\\
&  +g_{n }e^{x_{n }}\Big\{N_e(\sum^{n_0}_{n'=n+1}C_{nn'}+C_{nc})
 +\sum^{n-1}_{n'=1}A_{nn'}\{n'n\}+4\pi\int^{\infty}_{\nu_n} a_n(\nu)
                      \frac{J_{\nu c}}{h\nu} d\nu\Big\}\Big] \nonumber\\
& +\sum^{n_0}_{n'=n+1} b_{n'}\Big\{g_{n }e^{x_{n }}(N_eC_{nn'})
                    +g_{n'}e^{x_{n'}}A_{n'n}\{nn'\}\Big\}
  =-2\Big(\frac{2\pi m_ekT}{h^2}\Big)^\frac{3}{2}\alpha_{n} \nonumber\\
& -g_{n }e^{x_{n }}\Big\{N_eC_{nc}+4\pi\int^{\infty}_{\nu_n}
  a_n(\nu)\frac{J_{\nu c}}{h\nu}\exp\Big(-\frac{h\nu}{kT}\Big)d\nu\Big\}
  \qquad (n=1,2,\cdots,n_0)\ ,
\end{eqnarray}
where $\{nn'\}$ is the net radiative bracket for H$_{n,n'}$ line defined in
\citet{Thomas1961} and is given by
\begin{equation}\label{NetRadiativeBracket}
\{nn'\}=1-\displaystyle\frac{(N_{n}B_{nn'}-N_{n'}B_{n'n})J_{nn'}}
                            {N_{n'}A_{n'n}} \qquad (n<n').
\end{equation}

\bigskip

\subsection{Single-Flight Escape Probability}
We approximate the net radiative bracket by a single-flight escape probability
(e.g., \cite{Kastner1994}) for easier treatment of the line radiative process.
This approximation is introduced as an extension of Sobolev approximation for the
disk in which the velocity gradient is small. The single-flight escape probability
$\beta_{nn'}$ is defined as
\begin{equation}\label{betaformula}
 \beta_{nn'}=\displaystyle\frac{1}{4\pi}\displaystyle\int_{4\pi}
                     d\Omega \int^\infty_{-\infty} d\nu\psi (\nu-\nu_{nn'})
                     \exp\{-\tau^L_{nn'}(\nu)\}\ ,
\end{equation}
where $\Omega$ is the solid angle, $\nu_{n n'}$ is the central frequency of
H$_{nn'}$ line, and $\tau^L_{nn'}(\nu)$ is the optical thickness from a point in
the disk to the disk boundary for H$_{nn'}$ line at the frequency $\nu$ along the
path, for which we must take into account the velocity field. The function $\psi$
is an intrinsic broadening function and we assume thermal Doppler broadening.
Then, $\psi$ is given by
\begin{equation}\label{profilefunc}
 \psi(\nu-\nu_{nn'})=\displaystyle\frac{1}{\sqrt{\pi}\Delta\nu_D}
 \exp\Big\{-\Big(\frac{\nu-\nu_{nn'}}{\Delta\nu_D}\Big)^{2}\Big\},
\end{equation}
where $\Delta\nu_D$ is the thermal Doppler width, defined as 
\begin{equation}
 \Delta\nu_D=\displaystyle\frac{\nu_{nn'}}{c}{}
             \sqrt{\frac{2kT_{\rm disk}}{m_{\rm H}}}.
\end{equation} 

\bigskip

\subsection{Continuous Radiation}
The continuum mean intensity $J_{\nu c}$ is the sum of the mean intensity of
the direct and diffuse radiation:
\begin{equation}
 \label{AllContLight}
 J_{\nu c}=J^*_{\nu c}+J^d_{\nu c}.
\end{equation}
The direct radiation from the star, $J^*_{\nu c}$, is given by
\begin{equation}\label{DirectLight}
 J^*_{\nu c}=\frac{1}{4\pi}\displaystyle\int_{\Omega_*} I_\nu^*
 e^{-\tau_\nu^c}d\Omega\,,
\end{equation}
where $I_\nu^*$, $\tau^c_\nu$, $\Omega_*$ are the stellar radiation intensity
at the stellar surface, the continuum optical depth at the frequency $\nu$, and
the solid angle subtended by the star, respectively. When $\tau^c_\nu \simeq 0$,
$J^*_{\nu c}$ is approximated by
\begin{equation}\label{DirectLight1}
 J^*_{\nu c} \simeq W I_\nu^*\,,
\end{equation}
where $W$ is a geometrical dilution factor expressed by
\begin{equation}
 W=\frac{1}{2}\Biggl(1-\sqrt{1-\frac{1}{R^2}}\Biggr). 
\end{equation}
Here $R$ is the distance from the center of the star in units of $R_*$.

The diffuse radiation, $J^d_{\nu c}$, is expressed as 
\begin{equation}\label{DiffuseFormula}
 J^d_{\nu c}=\frac{1}{4\pi}\displaystyle\int_{4\pi} I^d_{\nu c}(s) d\Omega,
\end{equation}
where $I_{\nu c}^d(s)$ is the diffuse radiation intensity towards the
direction $s$, which is given by the formal integral of the radiative transfer
equation as
\begin{equation}\label{Transfer}
 I^d_{\nu c}(s)=\displaystyle\int_0^{\tau_{\nu b}^c} S_\nu e^{-\tau_\nu^c}d\tau_\nu^c.
\end{equation}
Here, $S_\nu$ is the source function and  $\tau_{\nu b}^c$ is the optical thickness
from a point in the disk to the disk boundary. Although electron scattering yields
the observed polarization, we do not include this effect in our calculations because
it does not affect the energy conversion in the disk and because the scattering
process makes the convergence in the iteration slower.

\bigskip

\subsection{Physical Constants}
$a_n(\nu)$ and $\alpha_n$ are adopted from \citet{Burgess1964} for $n=2s$ and $2p$.
We adopt \citet{Mihalas} and \citet{Johnson1972} for the other $a_n(\nu)$ and
$\alpha_n$, respectively. The gaunt factors for the bound-free and free-free
processes are taken from \citet{Kurucz1970}. The free-free absorption coefficient
$k_{\rm ff}(\nu)$ and the continuous emission coefficient $j_\nu$ are adopted from
\citet{Mihalas} and \citet{Hulst1962}, respectively. $A_{n'n}$ for $n=2s$ or $2p$
are from \citet{Karzas1961}, while the other $A_{n'n}$ values are from \citet{Wiese1966}.
$C_{nn'}$ and $C_{nc}$ are obtained from equations (36) and (39) in \citet{Johnson1972} 
for $n, n' \neq 2s, 2p$. $C_{2s,2p}$ is computed from equations (3) and (55) in
\citet{Seaton1955}. $C_{1,2s}$ is obtained through the interpolation of table X in
\citet{Callaway1985}. Because $C_{2s,3}$ and $C_{2p,3}$ are the sum of $C_{2s,3s}$,
$C_{2s,3p}$, $C_{2s,3d}$, and of $C_{2p,3s}$, $C_{2p,3p}$, $C_{2p,3d}$, respectively,
these six coefficients are computed through the interpolation of table III in
\citet{Callaway1987}. The coefficients $C_{2s,n'}$ and $C_{2p,n'}$ for $n'\geq4$ are
estimated from the collision rate of \citet{Johnson1972}, in the same manner as in
\citet{Krolik1978}. For $C_{2s,c}$ and $C_{2p,c}$, the assumption by \citet{Drake1980}
is adopted, i.e., $C_{nl,c}=C_{nc}$.

\bigskip

\subsection{Computation}
\subsubsection{Computational Scheme}
By assuming trial values for $b_n$, the coefficients for $b_n$ in equations
(\ref{ioniexcieq}) are obtained through integrations for the escape probability
defined by equation (\ref{betaformula}) and for the mean intensities of continuous
radiation. Therefore, the calculations of equations (\ref{betaformula}), 
(\ref{DirectLight}), and (\ref{DiffuseFormula}) and the solutions of equations 
(\ref{ioniexcieq}) and (\ref{NeDerive}) constitute an iteration loop in this order. 
We solve the transfer equation for the continuous radiation by this simple integral 
form, though it is known that the convergency is very slow when the disk becomes
optically thick. In actual calculations,
we assume the uppermost bound level, $n_0=10$, while we adopt $n_0=5$ when the
energy balance is discussed in section \ref{EnergyGainLoss}. For model 1, the
initial values were set to $b_n = 1/W$. Then, the results of model 1 were adopted
as the initial values for model 2, and so on. 

\bigskip

\subsubsection{Grid points and Interpolation}\label{grids}
The vertical boundary of the disk is set to the surface with
$N(\varpi,z)=10^4\,{\rm cm}^{-3}$, while the disk radius, $\varpi_{\rm disk}$, is
cut off at 95\% of the radius with $N(\varpi,0)=10^4\,{\rm cm}^{-3}$. The grid
points for calculating the $b_n$ factors are set denser near the star and the
equator. The fourteen grids are set exponentially for $\varpi$, while nine grids 
are set parabolically for $z$ ($z\geq 0$), including the boundary. Thus, the total
number of the grids is 126. We set the grid points in the following form 
\begin{equation}\label{varpiMesh}
 \varpi_i = \varpi_{\rm disk}^{\frac{i-1}{13}}\ (i=1,2,\cdots,14),\  {\rm and}\ 
 z_{i,j}  = z_{i,8}\left(\displaystyle\frac{j}{8}\right)^2\ (j=0,1,\cdots,8)\ ,
\end{equation}
where $z_{i,8}$ is the vertical boundary with $N(\varpi_i,z)=10^4\,{\rm cm}^{-3}$. 
Hereafter, we call these grid points as {\it population grid points}. Analytic
setting of the grids is convenient for the interpolation in the $(i,j)$ space with
an equal interval. Interpolation is made only for the $b_n$ factors, since all the 
relevant quantities are calculable through the $b_n$ factors in the isothermal disk. 
Actually, the $\log b_n$ values are interpolated. Easy and stable interpolation is
required for precise integrations described below.

\bigskip

\subsubsection{Integration}
Equations (\ref{ioniexcieq}) include many integrations in the coefficients of $b_n$. 
It is evident that the iteration converges to the false values if the integrations
appeared in equations (\ref{ioniexcieq}) are inaccurate. The 1\% accuracy of the
integration is aimed in our iterative calculations and also in calculations of the
emergent flux and the line profile. We checked the accuracy for several test
integrands in each integrations appeared in equations (\ref{ioniexcieq}). Typical
test integrands are based on the assumption of $b_n=1/W$. The integration is made
mostly by Simpson's 3/8 formula. 

The optical depth $\tau^L_{nn'}(\nu)$ in equation (\ref{betaformula}) is calculated
with the meshpoints concentrated around the place where the relative velocity is
zero along the path. The meshpoints for frequency integrals are set with an interval
of $0.08\Delta\nu_D$ for the range of $-5\Delta\nu_D$ to $5\Delta\nu_D$. The angular
integration in equation (\ref{betaformula}) is performed on a polar coordinate with
a polar axis parallel to the rotational axis of the star. The meridional integraion
is performed with 25 meshpoints, which distribute denser near the equator, and 24
meshpoints are set for the azimuthal integration with an equal interval.

The optical depth $\tau_\nu^c$ in equation (\ref{DirectLight}) is calculated with
120 meshpoints concentrated in the regions where the function $N^2(\varpi,z)/W$ is
large. The angular integration is performed by introducing a polar coordinate whose
polar axis passes the center of the star. We adopt 18 meshpoints from 0 to $\theta_*$ 
(the angular radius of the star) for the meridional integration, and 36 meshpoints
from 0 to $2\pi$ for the azimuthal integration. 

Since the continuous emission produced in the denser part near the star is effective
for the diffuse radiation, $J^d_{\nu c}$, in equation (\ref{DiffuseFormula}), we
introduce a polar coordinate whose polar axis passes the footpoint, $(\varpi, z)=(1,0)$ 
for the angular integration. The angular meshpoints are set denser near the footpoint
and towards the equatorial plane, depending on the population grid point. The
meridional integration is performed first, and then the azimuthal one. In the case of
the population grid point distant from the star, about 100 meshpoints are required for
the meridional integrals, while 12 meshpoints is enough to satisfy the 1\% precision
when the population grid point is located in the dense part, due to its isotopic
nature. For the azimuthal integration, 26 meshpoints are set, which are distributed
denser near the equator. 

The path integral in equation (\ref{Transfer}) is made by using Simpson's $3/8$ rule
for the calculation of the optical depth, while the trapezoidal rule is applied for
the integration of the source function. The segments for the calculation of the
optical depth are divided repeatedly until they have satisfied the constraint that the
absorption coefficient ratio for both edges of the segment is between 1/1.2 and 1.2
for all frequency meshpoints. Moreover, the additional constraint is added for the
optical depth of the segment, $\Delta\tau_\nu$,  at frequency $\nu$: the segment is
divided until satisfying $\Delta\tau_\nu<0.2$ for $0<\tau_\nu<2$, $\Delta\tau_\nu<2$
for $2<\tau_\nu<5$, $\Delta\tau_\nu<5$ for $5<\tau_\nu<10$, and we neglect the
contribution from the more distant region with $\tau_\nu >10$. As a result, the
number of the meshpoints along the path which passes only the optically thin
region is several tens, while it reaches several thousands when the path includes
the optically thick region.

Totally sixty four frequency meshpoints are set for the integration over the frequency 
to calculate the ionization rate in equation (\ref{ioniexcieq}). They consist of 13
meshpoints for the Lyman continuum, 10 meshpoints for the Balmer continuum, 7
meshpoints for the Pashen, Brackett, and Pfund continua, and 4 points for the other 
continua with $n>5$.

\bigskip

\subsubsection{Convergency}
Our simple iteration scheme requires many iterations in the case of optically thick
disk. In such a case, the convergence is slowest for $b_1$ at the place slightly
distant from the star around the equatorial plane where the Lyman optical thickness
towards the disk boundary is very large. When the ratio of the new $b_1$ value to its
previous one keeps monotonous variation during five consecutive iterations, the five
power of the ratio is multiplied to the new $b_1$ factor in order to accelerate the
convergence, while the root of previous and new $b_n$ factors is adopted as a next
trial value in order to relax the solution. 

By examining the convergence behavior of $b_n$, we found that all the $b_n$ factors
except $b_1$ in the dense region converge when all the ratios of newly calculated
departure factors $b_1$ to the previous factors are within $1\pm0.01$ at all the
population grid points. Such a behavior in iteration is understandable since our
models are not optically thick towards the vertical direction in the Balmer and the
other continua with longer wavelength, and since the fully ionized state is assured
from the assumed high disk temperature. As will be shown in section
\ref{Energy Flow and Conversion}, Lyman continuum does not play so important role in
the radiation field in the disk. Then, we decided to stop the iteration when the
above 1\% criterion is fulfilled for all $b_n$ values. When we compare such a result
with the results from further iteration and from the iteration with the initial $b_1$
values in opposite sense, a few tens percents deviation at most from the guessed
convergent values was noticed for $b_1$ in the dense region of model 7.

\bigskip

\subsubsection{Computational Time}
All calculations in this paper were made with the computer with the CPU of the AMD
Opteron 250 (2.4 GHz) at the Astronomical Data Analysis Center, ADAC, of the
National Astronomical Observatory of Japan. The calculation times for one iteration
are about twelve hours for the optically thin disk, about eighteen hours for 
the optically thick disk. The 1\% criterion was achieved after ten or less iterations 
for the optically thin case, and after about forty iterations for the optically thick
case. The total time is approximately several days for the optically thin case and
two weeks for the optically thick case. 

\bigskip

\section{Global Features}\label{NonLTEstate}
\subsection{Non-LTE State}
In order to show the optical property of our model sequence, we list, in table
\ref{ContTauTable}, the respective continuum optical thicknesses, $\tau_r, \tau_z$,
and $\tau_\phi$ at each continuum limit from the footpoint ($\varpi=1, z=0$) to the
disk boundary towards the radial, vertical, and azimuthal directions for our models
1, 4, and 7. Reminding an observational fact that the second Balmer jump due to the
strong Balmer shell lines appears in strong shell stars \citep{Divan1979}, we cover
the range of the mass loss rate for which the radial optical thickness at the Balmer
shortward edge is larger than unity. Model 1 is optically thin, while model 7 is
optically thick at all shortward edges. In the wavelength where the optical depth for
the continuum is large, the role of diffuse radiation becomes important in the
non-LTE state of the disk. In model 1, the contributions of the direct and diffuse
radiation to the ionization from the ground level are comparable in the vicinity of
the star, while the direct stellar radiation overwhelms in the longer wavelength
region. In model 7, the diffuse radiation overwhelms the direct radiation in the
dense part of the disk. Taking into consideration that our B1V star emits the
radiation in the Lyman and Balmer continua by 0.2\% and 93.3\% of the total luminosity,
respectively, we classify the optical property of the disk by the optical thickness
for the Balmer continuum. We regard that models 1 and 7 are the respective typical
cases for the optically thin and thick disks. Notice that the optical thickness at
the shortward edges of Balmer and Paschen series has the smallest values among those
at all the shortward edges. These facts imply that the stellar Balmer continuum
principally governs the radiation field in Be stars. We also note that an inequality
$\tau_z <\tau_r <\tau_\phi$ holds not only at the footpoint but also in the main
part along the equator. 

\begin{table}
 \begin{center}
 \caption{Continuum optical thickness towards the radial direction ($\tau_r$),
          the vertical direction ($\tau_z$) and the azimuthal direction
          ($\tau_\phi$) from the footpoint to the boundary in our three models. 
          All the bound-free processes and the free-free process are included. 
          The first row corresponds to the threshold wavelength: the integer 
          means the bound level related to the bound-free transition, while 
          the '$-$' and '$+$' signs indicate the shortward and the longward of the
          threshold, respectively. For example, '1c$-$' means the shortward edge
          of the Lyman limit. a(b) means $a\times 10^{b}$.}
 \label{ContTauTable}
  \begin{tabular}{lccccccccc}\hline
   & \multicolumn{3}{c}{model 1} & \multicolumn{3}{c}{model 4}
   & \multicolumn{3}{c}{model 7}\\[-2pt]\cline{2-10}
   & $\tau_r$ & $\tau_z$ & $\tau_\phi$ & $\tau_r$ & $\tau_z$ & $\tau_\phi$
   & $\tau_r$ & $\tau_z$ & $\tau_\phi$\\\hline
   1c$-$ & 5.75($+$0) & 3.52($-$1) & 1.26($+$1) & 4.65($+$2) & 2.45($+$1) &
           9.25($+$2) & 1.51($+$4)\footnotemark[$\dagger$] &
            2.41($+$3)\footnotemark[$\dagger$] & 4.85($+$4)\footnotemark[$\dagger$] \\
   1c$+$ & 4.68($-$4) & 3.01($-$5) & 1.05($-$3) & 2.59($-$2) & 2.24($-$3) &
           6.50($-$2) & 1.88($+$0) & 2.40($-$1) & 5.68($+$0)\\
   2c$-$ & 2.53($-$2) & 1.56($-$3) & 5.61($-$2) & 1.36($+$0) & 1.13($-$1) &
           3.35($+$0) & 9.64($+$1) & 1.21($+$1) & 2.89($+$2)\\
   2c$+$ & 2.29($-$3) & 2.49($-$4) & 6.56($-$3) & 1.85($-$1) & 2.38($-$2) &
           5.52($-$1) & 1.61($+$1) & 2.38($+$0) & 5.16($+$1)\\
   3c$-$ & 1.98($-$2) & 1.98($-$3) & 5.52($-$2) & 1.51($+$0) & 1.87($-$1) &
           4.43($+$0) & 1.26($+$2) & 1.87($+$1) & 4.05($+$2)\\
   3c$+$ & 7.54($-$3) & 1.02($-$3) & 2.37($-$2) & 7.12($-$1) & 1.02($-$1) &
           2.25($+$0) & 6.87($+$1) & 1.02($+$1) & 2.21($+$2)\\
   4c$-$ & 3.04($-$2) & 3.99($-$3) & 9.46($-$2) & 2.82($+$0) & 4.00($-$1) &
           8.84($+$0) & 2.68($+$2) & 3.99($+$1) & 8.65($+$2)\\
   4c$+$ & 2.11($-$2) & 3.06($-$3) & 6.80($-$2) & 2.10($+$0) & 3.09($-$1) &
           6.72($+$0) & 2.07($+$2) & 3.08($+$1) & 6.68($+$2)\\
   5c$-$ & 5.88($-$2) & 8.44($-$3) & 1.89($-$1) & 5.82($+$0) & 8.53($-$1) &
           1.86($+$1) & 5.73($+$2) & 8.52($+$1) & 1.85($+$3)\\
   5c$+$ & 5.05($-$2) & 7.45($-$3) & 1.63($-$1) & 5.06($+$0) & 7.49($-$1) &
           1.63($+$1) & 5.03($+$2) & 7.49($+$1) & 1.62($+$3)\\\hline
   \multicolumn{10}{@{}l@{}}{\hbox to 0pt{\parbox{180mm}
    {\par\noindent\footnotemark[$\dagger$] uncertain.}\hss}}
  \end{tabular}
 \end{center}
\end{table}

Figure \ref{bnFigure} shows the variation of the $b_n$ factors along the equatorial
plane as a function of the distance for our models 1, 4, and 7. The geometric
dilution factor $W$ is also shown with the dashed lines. The $b_n$ factors once
increase outwards from $b_n\sim 1$ in all our models, because the available stellar
radiation decreases outwards. The overpopulation is less remarkable in the higher
levels, since the collisional process operates more effectively in the higher levels.
The $1s$ and $2s$ departure factors increase towards the outer boundary, because
they are the ground level or the metastable level. It is noteworthy that the second
level population approaches to $1/W$, but does not exceed $1/W$, reflecting a fact
that the stellar Balmer continuum is a principal source of ionization. While, the
$b_n$ factors of the excited levels ($n=2p,3,4\ldots$) turn to decrease at some
distance, and finally converge to the nebular case A. Similar behavior was concluded
by \citet{Marl1978} (see figure 2 of \authorcite{Hirata1984} \yearcite{Hirata1984}).
We confirmed that the turning point with maximum $b_n$ reflects the change from
optically thick nature to thin nature in H$_{n-1,n}$ line, i.e., the leading member
of the series, thus degrading the $n$-th level population. The turning point shifts
outwards in the denser disk with larger mass loss rate, caused by optically thicker
nature. In the outer region, the disk becomes optically thin for the lines, first,
from the higher series members, then, to the lower series members. In the optically
thick disk, the LTE region is added in the innermost part where it is optically thick
in all wavelength regions. Figure \ref{Nbmap} shows the iso-contour map of the
departure factors $b_n$ on the meridional plane for our standard model 7 with $n_0=5$.
The $b_n$ factors become smaller as leaving from the equator, reflecting the smaller
optical thickness and subsequent more important role of the direct stellar radiation
from the polar region. For comparison, we also show the $W^{-1}$ contours for some
typical values. 

\begin{figure}
 \caption{$b_n$ factor variation on the equatorial plane for the models 1, 4, and 7.
          The dashed curves indicate $1/W$.} 
\label{bnFigure}
\end{figure}

\begin{figure}
 \caption{Contour map of the departure factor $b_n$, where $n=1,2s,2p,3,4,5$ on the
          meridional plane for model 7. The $\times$ signs indicate the position of
          the population grid points in our computation. The iso-contours of the
          dilution factor $W$ are also shown in each panel.}
 \label{Nbmap}
\end{figure}

\subsection{Energy Balance}\label{EnergyGainLoss}
Next, we calculate the radiation energy gain and loss in our isothermal disk and
examine the deviation from the radiative equilibrium. The calculations were made for
$n_0=5$. The energy loss at any point in the disk is given by
\begin{equation}
 E_{\rm loss}=4\pi\int j_\nu d\nu + \sum^{n_0}_{n'=2}\sum^{n'-1}_{n=1}
                h\nu_{nn'}N_{n'}A_{n'n}\beta_{nn'},
\end{equation}
where the first and second terms correspond to the loss through the continuous
process and the line process, respectively. The energy gain is expressed as
\begin{equation}
 E_{\rm gain}=4\pi\int k_\nu J_{\nu c}d\nu.
\end{equation}

If the energy loss is larger than the energy gain, the disk temperature should
become lower than the assumed one when the radiative equilibrium holds. We define
the ratio of $E_{\rm loss}$ to $E_{\rm gain}$ by $R_{\rm lg}$. Figure \ref{EnergyGL}
shows the ratio, $R_{\rm lg}$, along the equatorial plane for models 1, 7, and 9.
The model 9 ($T_{\rm disk}=12000$\,K) is shown for illustrating the case of the
lower disk temperature. The label T represents $R_{\rm lg}$, while the labels C
and L indicate the continuum and line contributions, respectively. 

In the case of optically thin disk (model 1), our isothermal assumption with 
$T_{\rm disk}=2/3 T_{\rm eff}$ fills almost the radiative equilibrium on the
equatorial plane. In model 7, however, the energy loss is larger than the energy
gain in almost all part of the equator. There exist two regions where the value
of $R_{\rm lg}$ reaches maximum: the inner part around $4R_*$ and the outer part
around $60R_*$. The formation of the inner cool region is caused by the ultraviolet
radiation deficiency due to the optically thick nature. The outer cool region is
formed because Lyman $\alpha$ becomes optically thin, thus the L$\alpha$ cooling
becomes effective. Outside of the L$\alpha$ cooling region, the $b_{2p}$ factor
drops drastically, so the L$\alpha$ cooling becomes ineffective. The energy loss
by the line process is negligibly small in the inner part, while it reaches a half
of the total loss in the outer L$\alpha$ cooling region. This L$\alpha$ cooling
region will not be observed in the optical spectral region, because no contribution
is expected due to its low density. In model 9 ($T_{\rm disk}=12000$\,K), the global
feature resembles to that in model 7. Note that the disk temperature in the innermost
LTE region should be higher than the assumed one, in the sense that it approaches
to the stellar effective temperature.

Figure \ref{EnerGL} shows the contour maps of $R_{\rm lg}$ on the meridional plane
for models 1, 7, and 9. One can see that the cooler region is concentrated to the
equatorial plane and the disk is hotter in general as $|z|$ becomes larger, caused
by the penetration of the stellar ultraviolet radiation.

\begin{figure}
 \caption{Ratio of the energy loss to the energy gain along the equatorial plane.
          The label T represents the ratio for the total energy loss $R_{\rm lg}$, 
          while the labels C and L indicate the continuum and line contributions, 
          respectively.}
 \label{EnergyGL}
\end{figure}

\begin{figure}
 \caption{Ratio of the energy gain rate to the loss rate on the meridional plane
          in models 1, 7, and 9. The upper three panels display the region within 
          $\varpi<20$, while the lower panels include more distant region up to 
          $\varpi<170$.}
 \label{EnerGL}
\end{figure}

\bigskip

\subsection{Energy Flow and Conversion}\label{Energy Flow and Conversion}
Let us discuss the radiation flow and conversion in the disk. Since the continuum
optical thickness is smallest towards the vertical direction (see table
\ref{ContTauTable}), the continuous radiation emitted in the disk tends to escape 
from the upper and lower surfaces of the disk. The inclination-angle dependency of 
the emergent flux will be given in the next section. 

Although the L$\alpha$ cooling is effective in the outer part of the disk, it is
found that the total amount of energy emitted in the lines from the whole disk is
negligible, compared with that in the continuum. This is because the amount of the
energy emitted in the region far from the star is much smaller than that near the
star. Hence, we consider only the continuous radiation in the following. We define
the continuum luminosity in Lyman, Balmer, \,$\cdots$ continua 
($L_1$, $L_2$,\,$\cdots; n=1-6$) for $n_0=5$ by 
\begin{equation}
 \label{Brightness}
  L_n=4\pi\int_0^\frac{\pi}{2}\sin idi
     \int^{\nu_{n-1}}_{\nu_n}  F_\nu(i)d\nu,
\end{equation}
\noindent where $F_\nu(i)$ is the flux integrated over the disk and the star towards
the inclination angle $i$ at the frequency $\nu$, $\nu_n$ being the threshold
frequency of each continuum band with the exceptions of $\nu_0=\infty$ and $\nu_6=0$.
We also define the similar quantity $L_{n}^{*}$ for the stellar radiation.

In figure \ref{EnergyFraction}, we show, for models 1, 7, and 9, the difference 
$L_{n} - L_{n}^{*}$ at each band, as well as the difference of the total luminosity, 
$L - L_*$ (designated as "all"), all being normalized by the stellar bolometric
luminosity $L_*$. The total luminosity of model 1 is slightly less than the stellar
luminosity, indicating the slightly higher {\it isothermal} equilibrium temperature. 
Total luminosity of model 7 is larger than the stellar luminosity, while that of 
model 9 is less than the stellar luminosity. This indicates that the isothermal 
equilibrium temperature is in between 12000~K and 16000~K for model 7. This
conclusion is consistent with the results in the radiative equilibrium models
calculated by \citet{Millar1999}, \citet{Carciofi2006}, and \citet{Sigut2007}, since 
they all gave $T_{\rm disk}$ around $0.6T_{\rm eff}$ as an equivalent isothermal disk 
temperature. 

Next, we discuss the radiation energy conversion in the disk. It is well known that 
stellar ultraviolet radiation is converted into radiation with longer wavelength and 
line radiation in the Be star disk. Figure \ref{EnergyFraction} 
clearly shows that Balmer continuum photons are converted into photons with longer 
wavelength. Similar conclusion was reported by \citet{Millar1999}. Stellar Lyman
continuum plays little role in the disk except the ground level regulation in the
outer envelope of the disk where the optical depth for Lyman continuum towards the
central star is small. In the optically thick disk, only a small fraction of the
Lyman continuum is converted into the optical and infrared radiation (models 7 and 9).
While, in the optically thin disk (model 1), Lyman continuum photons are even
produced in the disk from the Balmer continuum photons. This result comes from a
fact that a B1V star emits almost all the radiation not in the Lyman continuum but in
the Balmer continuum. This is a general conclusion for Be stars, because the stellar
Lyman continuum is more deficient in Be stars with later spectral types. 

\begin{figure}
 \caption{Disk luminosity in each band relative to the stellar bolometric
          luminosity for models 1, 7, and 9. The abscissa expresses each 
          band, e.g., 1c=Lyman continuum and 'all' indicates the total luminosity.
          The scale of the ordinate in the upper and lower two panels are different.} 
 \label{EnergyFraction}
\end{figure}

\bigskip

\section{Continuum Property}\label{ContProperty}
\subsection{Spectral Energy Distribution}
We describe several observable quantities for our models with $T_{\rm disk}=16000$~K
in present and next sections. Figure \ref{ContRadi} shows the spectral energy
distribution (SED) for models 1, 4, and 7. The SEDs for different inclination angles
are shown, together with the stellar SED. The infrared excess develops as the mass
loss rate increases and as the inclination angle decreases. The Lyman continuum shows
the deficiency or excess, depending on the inclination angle in the case of optically
thick disk. In model 7, the disk continuum contributes to the SEDs even in the Balmer
and Paschen continua. 

\begin{figure}
 \caption{Spectral energy distribution for three models.}
 \label{ContRadi}
\end{figure}

\bigskip

\subsection{UBV Bands}\label{UBV}
The $UBV$ magnitudes were calculated by using the response functions for the
photometric filters given in \citet{Buser1978}. Figure \ref{Vmag} shows the $V$
magnitude variation as a function of the mass loss rate. The brightness in the $V$
band increases as the mass loss rate increases and the inclination angle decreases.
The brightness decrease in the $V$ band is not resulted and only slight
brightening in the $V$ band is obtained in the equator-on case in our present model
sequence ($T_{\rm disk}=16000$\,K). We note, however, that the $V$ band brightness
decrease by 0.2 mag is obtained for $i=90^{\circ}$ in model 9
($T_{\rm disk}=12000$\,K). The observed range of $\Delta V\sim 0.5$ mag is well
covered by the mass loss rate less than $10^{-10}~\MO\,{\rm yr}^{-1}$. 

Next, we examine the inclination angle dependency of the emergent flux in each band.
Figure \ref{UBVflux} shows the fluxes in the $UBV$ bands for model 7 as a function
of $\cos i$, to which the projected area is proportional. It is seen that the
emergent fluxes are, roughly speaking, proportional to $\cos i$, though they are
slightly concentrated towards the lower inclination angle. 

\begin{figure}
 \caption{$V$-band magnitude variation as a function of the mass loss rate for the
          model sequence with $T_{\rm disk}=16000$\,K. We also show the disk mass
          in the upper part.}
 \label{Vmag}
\end{figure}

\begin{figure}
 \caption{Flux variation as a function of $\cos i$ in the $UBV$ bands for model 7.} 
 \label{UBVflux}
\end{figure}

Figure \ref{UBVFigure} shows the variations on the color-magnitude (CM) and
color-color (CC) diagrams. The $B-V$ color becomes redder and the brightness in the
$V$ band becomes larger as the mass loss rate increases. The amount of variation is
larger for smaller inclination angle. It is interesting to note that the variation
on the CM diagram almost lies on the same locus, regardless the mass loss rate and
the inclination angle. The color-color plot in figure \ref{UBVFigure}b indicates
that the $U-B$ color becomes bluer for the low inclination angle and redder for the
high inclination angle, as the disk develops. We confirmed that no color change in
$U-B$ occurs at $i\sim70^\circ$. We note that our present results are consistent
with the observed variations on the CM and CC diagrams for early Be stars
(e.g.,\,\cite{Hirata1981}). 

\begin{figure}
 \caption{(a) Variation on the Color-Magnitude diagram.
          (b) Variation on the Color-Color diagram.}
 \label{UBVFigure}
\end{figure}

\bigskip

\subsection{Far-Infrared Region}\label{IRExcess}
The free-free transition is a dominant process in this wavelength region. Therefore, 
the emergent flux is rather free from the non-LTE state and reflects the density and
temperature structure of the disk more directly. Figure \ref{IRmag} shows the
magnitude variation at four IRAS wavelengths as a function of the mass loss rate.
The monochromatic magnitudes were calculated at 12, 25, 60, and 100$\mu$m, since the
SED in this wavelength region is not complex. The brightening in magnitude scale
increases linearly with the increase of the mass loss rate in the pole-on case.

The difference between $i=0^\circ$ and $90^\circ$ reaches 1.5$-$1.8 mag in the case
of the larger mass loss rate, resulting one order of magnitude difference in the
mass loss rate in our models. We suggest that the pole-on approximation in
\citet{Waters1986} is valid for $i\leq 30^{\circ}$ and the systematic deviation is
expected for the higher inclination angle. Similar result has been concluded by 
\citet{Carciofi2007}. We found such a correlation between the mass loss rate given
by \citet{Waters1987} and the projected rotational velocity $v\sin i$ given by
\citet{Cote1987}, as is shown in figure \ref{Waters}. There is a tendency that the
mass loss rate becomes larger for larger stellar projected rotational velocity
$v\sin i$. We also note that, in figure \ref{Waters}, the stars with large $v\sin i$
($\sim$400 km\,s$^{-1}$) have the mass loss rates one order of magnitude larger than
that with the smaller $v\sin i$ ($\sim$ 100 km\,s$^{-1}$) for all the spectral type
groups. Our rough estimate suggests that the mass loss rate based on our model is
$10^{-12}-10^{-10}~\MO\,{\rm yr}^{-1}$, three orders of magnitude smaller than
that of \citet{Waters1987}, reflecting the difference in the initial velocity 
$v_{\varpi}(\varpi=1)$. 

\begin{figure}
 \caption{Monochromatic magnitude change at four IRAS wavelengths 
          (12, 25, 60, and 100~$\mu$m).}
 \label{IRmag}
\end{figure}

\begin{figure}
 \caption{Mass loss rate given in \citet{Waters1987} against $v\sin i$. 
          Different spectral types are distinguished by different symbols.}
 \label{Waters}
\end{figure}

Figure {\ref{IRASflux} shows the flux variation at four IRAS wavelengths as a function
of $\cos i$ for model 7. Note that the disk contribution overwhelms the stellar flux
in the IRAS wavelength region. Again, the flux increases almost linearly with $\cos i$.
However, due to the opaque nature in this wavelength region, the disk has a finite
opaque area, thus resulting the infrared excess even in the equator-on view. Note
that the $\cos i$ dependence of the infrared flux is convex, in contrast to that in
the optical region (see figure \ref{UBVflux}).

\begin{figure}
 \caption{Flux variation as a function of $\cos i$ in the IRAS bands for model 7.} 
 \label{IRASflux}
\end{figure}
\bigskip

\section{Line Property}\label{LineProperty}
\subsection{Balmer Line Profiles}\label{Line}
Balmer line profiles were calculated under the assumption that the intrinsic
broadening function is given by the thermal motion to see the effect of velocity
field, although Stark broadening becomes important in the denser part of the optically
thick disk. Our models cover the large range of the continuous optical thickness at
H$\alpha$ from $\tau_r=0.01$ (model 1) to $\tau_r=375$ (model 8) (see also table
\ref{ContTauTable}). In the optically thick models, the contribution from the disk
continuum is noticeable even in the visual region, as is shown in the previous section.
This means that the resulting Balmer line profiles are complex, disturbed not only
by the shell absorption, that is, the line absorption of the photospheric radiation,
but also by the line absorption of the disk continuum. The latter is attributable to
the so-called {\it pseudo-photosphere} (\cite{Harmanec1994}, \yearcite{Harmanec2000}).

In figure \ref{Balmer} we show the rectified H$\alpha$, H$\beta$, and H$\gamma$ line
profiles in three models 1, 4 and 7 for various inclination angles. The photospheric
component is not included in this figure. In the case of $i\sim 0^{\circ}$, the narrow
double-peak profile appears. This special case will be discussed in section \ref{ps}.
The wine-bottle type profile is traced for $i=15^\circ$ and $30^\circ$ in model 7.
The broad double-peak profile develops for the larger inclination angle. The peak
separation is larger for the smaller mass loss rate and for the larger inclination
angle. The formation of the peak separation will be also discussed in section \ref{ps}. 
The strong shell component appears for $i=90^{\circ}$ in models 4 and 7. These
behaviors, including the narrow double-peak emission line profile in the pole-on case,
are similar to those obtained by \citet{Hummel1994}. We also note that the central
quasi-emission (CQE) is noticed for $i=90^{\circ}$ of model 1 (for CQE, see
\cite{Hanuschik1995}).

\begin{figure}
 \caption{H$\alpha$, H$\beta$, and H$\gamma$ line profiles for our three models. 
          The photospheric line is not considered in these profiles. The dotted,
          dashed and solid lines are the line profiles for models 1, 4 and 7,
          respectively. Note that the top panels have the vertical scales different
          from the other lower panels.} 
 \label{Balmer}
\end{figure}

The H$\alpha$ line profiles, in the flux form, with the inclusion of the photospheric 
component are illustrated in figure \ref{Halpha4} for model 4, where we also show
the contribution from various components: the emission component, the shell absorption
component, and the photospheric component. The emission component includes the line
and continuum originated in the disk. The intrinsic photospheric profile was taken
from ATLAS9 \citep{Kurucz1993} with the same parameters as for the stellar SED. This
profile is put on the rotating stellar surface with the critical equatorial rotational
velocity of 591 km\,s$^{-1}$ without limb darkening. One can see the {\it veiling}
effect by the disk continuum, which is more noticeable in the smaller inclination angle. 

For illustrating a case contaminated heavily by the disk continuum, we show, in 
figure \ref{Halpha7}, the H$\alpha$ profiles in model 7. In addition to the veiling
effect, the emission component has absorption features below the disk continuum in
the wing part for $i=30^{\circ} - 75^{\circ}$ and in the central part for $i=90^{\circ}$.
These absorption features are formed by the line absorption of the disk continuum,
i.e., in the pseudo-photosphere. This suggests that 1) for the intermediate inclination
angle, it is difficult to segregate the photospheric component from the observed
profile, caused by the broad shell absorption and the pseudo-photospheric absorption, 
and 2) for $i\sim 90^{\circ}$, the observed central intensity could be almost zero
even if the disk continuum is present. We note that the {\it emission} equivalent
width rectified by the disk continuum and integrated over the line feature may become
even {\it negative}, caused by the dominant pseudo-photospheric absorption feature.
This is notable in the higher Balmer lines in our models. We also note that a weak
central shell absorption feature appears for $i\geq 75^{\circ}$ in model 7. Our major
results on H$\alpha$ line are similar to those of \citet{Hummel1994} except the
veiling effect and the pseudo-photospheric absorption, which are newly introduced
in this paper.

\begin{figure}
 \caption{H$\alpha$ line profiles with the inclusion of the photospheric component 
          in model 4. The contribution of the photospheric line, the shell absorption
          and the disk emission component are also shown. The ordinate is in the flux
          form.}
 \label{Halpha4}
\end{figure}

\begin{figure}
 \caption{Same as figure \ref{Halpha4} but in model 7.}
 \label{Halpha7}
\end{figure}

\subsection{Peak Separation}\label{ps}
As stated in section \ref{Line}, the narrow double-peak emission profile appears in
the case of $i\sim 0^{\circ}$. The same characteristics was concluded and interpreted
by \citet{Hummel1994} in terms of the non-coherent scattering (NCS). Figure \ref{tau1}
illustrates the vertical change of the H$\alpha$ source function at the distance of
$\varpi=1.06$ in model 7. The optical depth at the line center is also shown. One can
see that the source function increases as the optical depth increases in the effective
line formation region. Similar tendency is kept in the inner part of the disk, where
the source function is large. Recalling that no velocity gradient exists in the
vertical direction in our model, the central reversal is naturally formed. 

In figure \ref{VRPeak}a, we show the resulting peak separation as a function of
$\sin i$. As $\sin i$ increases, the profile characteristic changes from double-peak 
profile caused by NCS, passing the single-peak profile, to the double-peak profile of 
rotation origin. The peak separation caused by NCS is within two or three Doppler
widths in our calculations and increases as the mass loss rate increases, caused by
the stronger central reversal. The domain of this type of double-peak profiles expands
further to the larger inclination angle as the mass loss rate increases. The same type
profiles are also formed in H$\beta$ and H$\gamma$ lines with the peak separations
narrower than those in H$\alpha$ line (see figure \ref{Balmer}), reflecting their
smaller optical thickness. 

The peak separation originated in the rotational velocity field may be broadened by 
the shell and the other absorption components. In figure \ref{VRPeak}a, we also
show the peak separations derived from the emission component only, the profile
including the shell absorption component, and the profile including further the
photospheric component for models 4 an 7. While, we show only the peak separation
derived from the emission component for model 1, because the the emission component
is very weak, so the determination of peak position is difficult when the shell
absorption and the photospheric component are taken into account. One can see that
the effect of making peak separation larger by such absorption features is small even
in the weak emission profiles for $i\sim 90^{\circ}$ in model 4. The peak separations
in model 7 is little influenced by the background absorption features, since the pure
emission profiles have sharper peaks. Note that the large deviation of the peak
separation at $i=90^{\circ}$ from those at the other inclination angles is originated
in the pure emission component itself.

\begin{figure}
 \caption{Vertical H$\alpha$ source function variation at $\varpi=1.06$ in model 7.
          Several optical depths at the line center are indicated.}
 \label{tau1}
\end{figure}

Now, let us examine the peak separation of the double-peak profile in the higher 
inclination angle. The peak separation of the double-peak emission line profile in
Be stars has been often used for estimating the outer radius of the disk.
Observationally, it is well known that the peak separation is smallest in H$\alpha$
and becomes larger for the higher Balmer lines, reflecting the emitting sizes in each
lines. The outer radius of the disk, $R_{\rm disk}$, in the Keplerian rotation law
is estimated as
\begin{equation}
 \frac{R_{\rm disk}}{R_*}=
 \bigg(\frac{2v_\phi(\varpi=1)\sin i}{\Delta V_{\rm peak}}\bigg)^{2},
 \label{KeplerRadius}
\end{equation}
where $\Delta V_{\rm peak}$ is the peak separation of the double peak profile.
This formula is derived from the profile calculation for the rotating flat disk
with a finite outer radius, under the assumption that the line surface brightness 
distribution is given (e.g.,\,\cite{Huang1972}, \yearcite{Huang1973}, \cite{Hirata1984}).
The velocity at the maximum intensity corresponds to that at which the 
iso-line-of-sight velocity curve contacts with the outer radius, unless the surface 
brightness distribution is a steeply decreasing function of the distance. If the
surface brightness is locally proportional to $\varpi^{-m}$, it is easy to show
that the region with $m>2.5$ does little contribute to the emission profile in the
Keplerian disk (see equations (4)$-$(7) in \cite{Hirata1984}). Then, the peak
separation in our model which extends infinitely towards the radial direction is
thought to reflect the maximum radius with $m<2.5$.

The peak separation given by equation (\ref{KeplerRadius}) is proportional to
$\sin i$ when the outer radius is fixed. While, in figure \ref{VRPeak}a, we notice
a quasi-linearity except for $i\sim 0^{\circ}$ and $i\sim 90^{\circ}$. Hence, the
calculated peak separation reflects somehow the effective disk radius, although the
quasi-linear part does not pass the zero point, in contrast to the implication from
equation (\ref{KeplerRadius}). It is possible to derive the effective outer radius
by fitting the slope in the quasi-linear part of figure \ref{VRPeak}a to that in
equation (\ref{KeplerRadius}). Then, we obtain $R_{\rm disk}=1.7,\ 4.9$ and $12.3~R_*$
for models 1, 4, and 7, respectively.

Although we did not calculate the line surface brightness for our models, it is
possible to examine how the peak position is formed in our models. Figure \ref{surface}
shows the variation of the escape probability $\beta$, the source function $S$, and
$p=N_{3}\sum_{n=2{\rm s},2{\rm p}} A(3,n)\beta(n,3)$ for H$\alpha$ line in model 7
along the equatorial plane. The last quantity is proportional to the H$\alpha$
photon energy which escapes from the equatorial plane to the outside of the disk and
may be regarded as the angle-averaged surface brightness distribution on the equatorial
plane. With a slope of $m=2.5$ for $p$, we obtain $R_{\rm disk}=$ 1.3, 3.5,
and 7.8~$R_*$ for models 1, 4, and 7, respectively. These values are, roughly speaking, 
consistent with those estimated from the slopes in figure \ref{VRPeak}a, mentioned
before. The source function illustrated in figure \ref{surface} essentially behaves
in the same manner as $p$ in the main part of the disk, having a slower decline in the
inner part and a steeper decline in the outer part. Hence, we can guess the formation
region of any lines from their source function behavior along the equatorial plane. 
The source function for H$_{nn'}$ line is approximately proportional to $b_{n'}/{b_n}$. 
From figure \ref{bnFigure}, it is evident that the higher Balmer line is formed in
the region nearer to the star than the region where the lower Balmer line is formed,
thus has a larger peak separation.

In figure \ref{VRPeak}b, we show the H$\alpha$ outer radii estimated from equation 
(\ref{KeplerRadius}), together with the radii estimated from $p$ and $\beta(2p,3)=0.5$
at the equatorial plane. We choose $\beta(2p,3)=0.5$ as the effective boundary between
the optically thick and thin regions at H$\alpha$. Roughly speaking, the radii estimated 
by equation (\ref{KeplerRadius}) fall in between our two rough estimates. However, 
the direct application of equation (\ref{KeplerRadius}) to the observation gives only 
a rough estimation of the disk radius. \citet{Hummel1994} obtained the same conclusion 
and discussed the peak separation formation, based on the emissivity behavior for his
$N(\varpi,0)\propto\varpi^{-3}$ models.

\begin{figure}
 \caption{(a) H$\alpha$ peak separation against $\sin i$ for three models. The peak
          separation obtained from the emission component, the emission+shell
          component, and the emission+shell+photospheric components are designated
          by "+", $\circ$, and $\times$, respectively.
          (b) Disk radius calculated from equation (\ref{KeplerRadius}). Symbols
          are the same as in (a). We also show two rough theoretical estimations:
          the distance where the escaping H$\alpha$ photons begins to decrease
          rapidly from the equatorial plane (thin dotted line with a designation
          of $m=2.5$) and the distance where the escape probability of the H$\alpha$
          photon becomes 0.5 at the equatorial plane (thick dashed line with a
          designation of $\beta=0.5$). See the text for more details.}
 \label{VRPeak}
\end{figure}

\begin{figure}
 \caption{Escape probability, source function, and 
          $p=N_{3}\sum_{n=2{\rm s},2{\rm p}} A(3,n)\beta(n,3)$ at the equator
          for H$\alpha$ line in model 7. See the text for the details.}
 \label{surface}
\end{figure}

\bigskip

\subsection{Equivalent Width and Balmer Decrement}
Physically meaningful line emission strength would be the total flux calculated from
the emission component above the disk continuum, illustrated in figures \ref{Halpha4}
and \ref{Halpha7}. We call it a line emission power, hereafter. However, the synthesized
emission profiles are more complex one, disturbed by the disk continuum, by the shell
component in the case of large inclination angle, and by the pseudo-photospheric
absorption in the optically thick disk. Here, we simply define the emission equivalent
width, $W_{\alpha}$, as the area above the apparent continuum level in figure
\ref{Balmer}. This corresponds approximately to the equivalent width above the
photospheric absorption profile in actual observation. 

Figure \ref{EWHa} shows the variation of $W_{\alpha}$ as a function of the mass loss
rate. The computed H$\alpha$ equivalent widths are rather small and are similar to
those reported by \citet{Hummel1994} and \citet{Hummel1997} for their models with 
$N(\varpi,0) \propto \varpi^{-3}$. They are 6\,\AA~at most and do not reproduce the
observed ones (e.g.,\,\cite{Briot1971}; \cite{Dachs1986}; \cite{Dachs1990},
\yearcite{Dachs1992}). \citet{Sigut2007} have also concluded that the H$\alpha$
luminosity for their models with $\rho \propto \varpi^{-3.5}$ is too faint when
compared with the observations. This problem will be discussed in the last section. 
At the high end of the mass loss rate larger than $5\times 10^{-11} M_{\odot}\,{\rm yr}^{-1}$, 
the equivalent width gradually decreases due to the veiling effect for
$i \leq 60^{\circ}$. In figure \ref{EWHa}, we also show the equivalent width
integrating over the profile, including the absorption feature in the case of
$i=90^{\circ}$ (dotted line) in order to demonstrate the effect of the absorption
feature: the broad shell absorption component overwhelms the emission component when
the mass loss rate is low (see figures \ref{Balmer} and \ref{Halpha4}). This implies
that the development of a broad absorption feature is the first spectroscopic evidence
of a new mass loss event in the case of $i\sim90^{\circ}$. 

Figure \ref{HalphaI} shows (a) the line emission power and (b) the equivalent width 
of H$\alpha$ against $\cos i$. The H$\alpha$ emission power is free from the veiling
effect and from the shell absorption by its definition. Declines at $i=90^{\circ}$
in models 4 and 7 are caused by the central pseudo-photospheric absorption in figure 
\ref{HalphaI}a and by the central shell and pseudo-photospheric absorption in figure 
\ref{HalphaI}b. The H$\alpha$ emission equivalent width in figure \ref{HalphaI}b
suffers strong veiling effect for $i \leq 60^{\circ}$ in model 7. From figure
\ref{HalphaI}a, we conclude that H$\alpha$ photons also tend to escape towards the
polar direction, in spite of the existence of velocity gradient. However, such a
tendency is less remarkable, compared with the continuous radiation (see figures
\ref{UBVflux}, \ref{IRASflux}, and \ref{HalphaI}a). We confirmed that H$\beta$ and
H$\gamma$ lines show the same tendency. 

\begin{figure}
 \caption{Variation of the H$\alpha$ emission equivalent width as a function of the
          mass loss rate. The disk mass is also shown in the upper part. The dotted
          line corresponds to the equivalent width with the inclusion of the absorption
          feature below the apparent continuum level.}
 \label{EWHa}
\end{figure}

\begin{figure}
 \caption{(a) Emission power, (b) emission equivalent width of H$\alpha$ as a function
          of $\cos i$.}
 \label{HalphaI}
\end{figure}

The Balmer decrement may be defined as the ratio of the fluxes above the apparent
continuum level in figures \ref{Halpha4} and \ref{Halpha7}, corresponding to our
definition of the emission equivalent width. Here, we call it {\it observable}
Balmer decrement. We also define {\it theoretical} Balmer decrement derived from
the line emission powers. Figure \ref{Decrement} shows (a) theoretical and (b)
observable Balmer decrements in the form of the ratio of H$\alpha$ to H$\beta$
($D_{34}$) versus the ratio of H$\gamma$ to H$\beta$ ($D_{54}$). For the observable
Balmer decrement, we do not adopt the cases in which $W_\alpha$ is less than 0.1\,\AA,
because the results are less accurate and are also meaningless from the observational
point of view. Several cases for high inclination angles are lost in figure
\ref{Decrement}b due to this cut-off in H$\gamma$ and H$\beta$ lines. For comparison,
we also show a nebular case B solution for $T_{\rm e}=10000$\,K \citep{Brocklehurst1971}
by a large plus marked "Case B". The theoretical Balmer decrements are located in the
left of case B, while the observable Balmer decrements extend towards right and lower
direction, caused by the shell absorption effect for high inclination angle (see the
cases for $i=75^{\circ}$). Our results in figure \ref{Decrement}b fall in the
observed area given in \citet{Dachs1990}. In both diagrams, the Balmer decrement is
steeper for higher inclination angle. This is consistent with the observational
implication of steeper decrement for shell stars suggested by \citet{Dachs1990}. 

\begin{figure}
 \caption{Balmer decrement for H$\alpha -$H$\gamma$ lines. D$_{34}$ is the ratio of
          H$\alpha$ to H$\beta$ and D$_{54}$ is the ratio of H$\gamma$ to H$\beta$.
          (a) {\it theoretical} diagram derived from the line emission power,
          (b) {\it observable} diagram derived from the equivalent width above
          the apparent continuum. See the text for the details. The nebular case B
          is also indicated by a large plus sign.}
 \label{Decrement}
\end{figure}

\bigskip

\section{Summary and Concluding Remarks}\label{Conclusion}
The non-LTE state of hydrogen atoms in the isothermal steady-state decretion disk 
around a B1V star has been calculated, based on the $\Lambda$ iteration for the
continuous radiation and the single-flight escape probability for the line radiation.
Thus, we obtained the continuous radiation field without approximation. The basic
physical process in our models was presented. We described the non-LTE state in
terms of the $b_n$ factors. We examined the radiation flow and radiation conversion
in the disk. The local and global deviations from the radiative equilibrium in our
iso-thermal models were examined through counting the energy gain and loss. We also
calculated several photometric and spectroscopic quantities and examined their
dependency on the mass loss rate and the inclination angle. Some implications to the
observable quantities were also presented.

\bigskip

Our main conclusions are summarized as follows:
\begin{enumerate}
\item The $b_n$ factors increase outwards from $b_n\sim 1$. The $b_n$ factors of
$2p$ and higher levels then turn to decrease at the distance where the H$_{n, n-1}$
line becomes optically thin and finally tend to a nebular case A solution. In
the optically thick disk, the LTE region is added in the innermost region of the disk.

\item Stellar Balmer radiation plays a definitive role in the radiation field of
the disk. The conversion of the stellar Balmer radiation into the longer wavelength
continuous radiation and the line emission is the main process in radiation field of
the Be star disk. Stellar Lyman continuum contributes only to the ionization in the
outer envelope of the disk where the disk is optically thin towards the central star.

\item The radiative equilibrium almost holds in the disk with
$T_{\rm disk}=2/3T_{\rm eff}$ in the optically thin case for the Balmer continuum. 
Similar conclusion is obtained by \citet{Carciofi2006}. When the disk becomes
optically thick, there appear two cool regions: the inner region caused by the
deficient Balmer continuous radiation and the outer region caused by the Lyman
$\alpha$ cooling. 

\item The optical thickness of the disk is smallest towards the vertical direction
for the continuum. Hence, the continuous radiation tends to escape from the disk
vertically. Our models with the mass loss rate less than $10^{-10} M_{\odot}\,{\rm yr}^{-1}$ 
cover well the observed variations in the $UBV$ and IRAS bands. The infrared excess
in the IRAS band for $90^\circ$ is smaller than that for $i=0^\circ$ by 1.5$-$1.8 mag.
We suggest that the pole-on approximation in \citet{Waters1986} is valid only for
$i\le 30^\circ$. The infrared mass loss rate based on our model is
$10^{-12}-10^{-10}~\MO\,{\rm yr}^{-1}$, three order of magnitude smaller than that
of \citet{Waters1987}.

\item General characteristics of the observed profiles of the classical Be stars
like the wine-bottle, double peak, and shell profiles are reproduced in our non-LTE
model, just as was shown by \citet{Hummel1994}. However, the isothermal steady-state
decretion disk does not give the observed large H$\alpha$ emission intensity.

\item The pseudo-photospheric absorption feature through the line absorption of the
disk continuum appears in the case of the large mass loss rate. 

\item The peak separation of the double-peak profile is almost determined from the
emission profile, that is, the shell absorption and the photospheric absorption
affect the peak position only a little bit. Though the peak separation gives a rough
measure of the outer radius of the disk, the resultant radius still depends on the
inclination angle. This is just the same conclusion given by \citet{Hummel1994}.
\end{enumerate}

Most serious result in our calculation based on the $N(\varpi,0)\propto\varpi^{-3.5}$
law is  that we did not get the large H$\alpha$ emission equivalent width observed
in early Be stars. As the mass loss rate increases, our isothermal model sequence
with $T_{\rm disk}=2/3T_{\rm eff}$ deviates more from the radiative equilibrium in
the sense that the radiation loss from the disk is larger than the energy input from
the star. This means that the radiation energy lost from the disk comes partially
from the compensation of the internal energy in the disk with the temperature higher
than the equilibrium temperature. Hence, the H$\alpha$ intensity will not increase 
even if we introduce the radiative equilibrium model with the same density structure.

It is highly probable that we should introduce some factors which enhance the line 
emission, compared with the continuous emission. The Be star disk may extend more
towards the vertical direction. We introduced the turbulence velocity which is equal
to the sound velocity. However, we obtained only a small enhancement in the H$\alpha$
equivalent width. Introduction of high velocity gradient may improve the situation.
We calculated the non-LTE state for the model proposed by \citet{Stee1994}, keeping
the disk mass of our standard disk model. However, the resultant H$\alpha$ equivalent
width {\it does} decrease drastically, since the high density region becomes compact.
\citet{Sigut2007} adopted the $N(\varpi,0) \propto \varpi^{-2.5}$ law and obtained
the H$\alpha$ luminosity comparable to the observed values. We also made the preliminary
calculation for the $N(\varpi,0)\propto\varpi^{-2.5}$ law with the same mass loss
rate as in model 7 and obtained the H$\alpha$ emission equivalent width of
80\,\AA~for $i=90^{\circ}$, reflecting the large projected area of the optically
thick region at H$\alpha$. This result may suggest that the density in actual Be
stars decreases more slowly than our model with $N(\varpi,0)\propto\varpi^{-3.5}$,
as suggested by \citet{Sigut2007}. This is possible if the disk temperature decreases
outwards, caused by deficient Balmer continuous radiation in the optically thick region.

Our adoption of the single-flight escape probability for the lines may not be valid
in the solution of the line radiative transfer. It is evident that this approximation 
underestimates the net radiative bracket, because the local escape of photons is
replaced by the global escape from the disk. It is also important to get the exact
solution numerically for the line transfer problem.

\bigskip

\section*{Acknowledgements}
\noindent The authors thank A.\,T.\,Okazaki who provided us his code for the
transonic decretion disk model. All of the calculation were carried out on the
computer system at the Astronomical Data Analysis Center, ADAC, of the National
Astronomical Observatory of Japan. The authors also thank to the anonymous referee
for many invaluable comments on the original manuscript, by which the presentation
of this article is much improved.

\end{document}